\documentclass[aps,twocolumn,nofootinbib]{revtex4}
\usepackage[latin1]{inputenc}
\usepackage{epsfig}
\usepackage{graphicx}
\newcommand{\beq}{\begin{equation}}
\newcommand{\eeq}{\end{equation}}

\begin{document}

\title{Epidemic spreading}

\author{Tânia Tomé and Mário J. de Oliveira}
\affiliation{Universidade de São Paulo, Instituto de Física,
Rua do Matão, 1371, 05508-090 São Paulo, SP, Brazil}

\begin{abstract}

We present an analysis of six deterministic models for
epidemic spreading. The evolution of the number of
individuals of each class is given by ordinary
differential equations of the first order in time, which are
set up by using the laws of mass action providing
the rates of the several processes that define each model.
The epidemic spreading is characterized by the frequency
of new cases, which is the number of individuals
that are becoming infected per unit time. It is also
characterized by the basic reproduction number, which
we show to be related to the largest eigenvalue of the
stability matrix associated with the disease-free solution
of the evolution equations. We also emphasize
the analogy between the outbreak of an epidemic with a critical phase 
transition. When the density of the population reaches
a critical value the spreading sets in, a result that
was advanced by Kermack and McKendrick in their study
of a model in which the recovered individuals acquire permanent
immunization, which is one of the models analyzed here.

\end{abstract}

\maketitle

\section{Introduction}

The control and possibly the eradication of an infectious disease 
is not fully successful without the knowledge of the mechanisms
that determine its spreading in a population.
A major contribution in this direction is provided by  
the theoretical study of epidemic spreading through the
use of deterministic and stochastic models.
The theoretical study of the epidemic spreading
\cite{bailey1957,anderson1991,renshaw1991,mollison1995,
hastings1997,keeling2008} is not new but it
did not properly start before the physical basis for the
causes of the infectious disease was established in the
second half of the nineteenth century \cite{bailey1957}.
By the end of the nineteenth century the mechanism of 
epidemic spreading, revealed by bacteriology, allowed the 
development of the first epidemic models \cite{bailey1957}.

Hammer in 1906, carried out an analysis of the measles epidemic
considering that the infection spreads from person to person
and depends on the number of the susceptible \cite{bailey1957}.
In his studies of the relation
between the number of mosquitoes and the incidence of malaria,
Ross starting from 1908 formulated models for the transmission
of the infectious diseases \cite{bailey1957}.
In his book on the prevention of malaria of 1911 \cite{ross1911}, 
he employed ordinary differential equations 
to describe the evolution of the number of affected and 
unaffected individuals by a disease, and wrote the equations 
in terms of rates of several types, pointing out the major role
of the infection rate.

A more general theory than that of Ross, but employing
similar ideas, was developed by Kermack and McKendrick in 1927
\cite{kermack1927}.
They proposed a model for an epidemic consisting of
recovered individuals having permanent immunization.
They were able to solve the differential equations 
that governs the time evolution of the number of
susceptible, infected, and recovered, arriving
at a remarkable theorem concerning the spreading threshold
of an epidemic \cite{bailey1957,kermack1927}.
If the density of the susceptible
is smaller than a certain value, the epidemic does not
outbreak. In other terms, if the infection rate is smaller
than a critical value the disease does not spread.

In 1929, Soper developed deterministic models for measles
by using difference equations \cite{bailey1957},
and by assuming that the operations
of transmission are analogous to the mass law action of chemistry
\cite{soper1929}. The equations developed by Kermack and
McKendrick were also in accordance with this law
which would become one of the most important concepts in 
theoretical epidemiology \cite{anderson1991}.

The deterministic models, employing ordinary differential equations,
were carried further mainly after around 1945 \cite{bailey1957}.
In 1952, Macdonald introduced a concept, which he called
basic reproduction rate, concerning the threshold of spreading,
with reference to the threshold theorem of Kermack
and McKendrick \cite{macdonald1952,heesterbeek2015}.
However, it was not before 1975 when the
concept was reintroduced by Hethcote and by Dietz,
that it became widely used \cite{heesterbeek2002}. 

Around 1945, the stochastic approach for epidemic spreading,
which had been originated earlier, received further
development by Bartlett \cite{bailey1957}.
He transformed the Kermack and McKendrick
epidemic model into a stochastic model by treating the
numbers of susceptible and infective individuals as stochastic
variables \cite{bartlett1949}. The corresponding master
equation in these two variables was obtained by Bailey
\cite{bailey1957,bailey1953} and the stochastic version of
the Kermack and McKendrick threshold theorem was advanced
by Whittle \cite{whittle1955}.
The model included two processes:
the recovery of infected individual, who acquires permanent
immunization, and the infection of a susceptible by the contact
with an infective individual. These stochastic models
can be understood as discrete generic random walk in a space
whose axes are the numbers of various classes of individuals
\cite{nisbet1982,tome2015L}.

A distinct type of stochastic models for epidemic spreading
takes into account the spatial structure in which real
infectious diseases spread in a population
\cite{harris1974,grassberger1983,grassberger1983a,
ohtsuki1986,satulovsky1994,durrett1995,
antal2001,dammer2003,souza2010,tome2011,souza2013,tome2015,ruziska2017}.
The spatial stochastic models are formulated in terms of
several stochastic variables, one for each
individual, and which take values corresponding to the
class an individual belongs in with respect to the disease
\cite{tome2015L}. Usually, they are defined on a lattice that represents
the spatial structure. 

The spatial stochastic model allows us to derive the stochastic
model of the generic random walk mentioned above by a suitable
reduction of stochastic variables. This reduction can be made at the first
level, represented by the numbers of individuals of each
class, or at the second level, which takes into account the
pairs of individuals \cite{tome2009,tome2011}.
It is also possible from the spatial
stochastic models to reach the deterministic evolution equations
at the first level, which are the usual evolution equations
such as the ones studied here, or at the second level where the
numbers of pairs of individuals are taken into account
\cite{satulovsky1994,souza2013,ana2017}.

Here we only consider the deterministic approach to
the epidemic spreading, which still is a relevant 
approach to the study of epidemic spreading
of various infectious diseases
\cite{adriana2006,burattini2008,pinho2010}. 
We analyze models that 
are appropriate for the direct transmission of the disease,
that is, when the transmission occurs from person to person,
which is the case of most common infectious diseases such
as measles, chickenpox, mumps, rubella, smallpox, common cold, and
influenza, as well as the present covid-19.
These models include the susceptible-infected-susceptible (SIS)
\cite{ross1911},
the susceptible-infected-recovered (SIR) \cite{kermack1927}, and
the susceptible-exposed-infected-recovered (SEIR) \cite{dietz1976}.
The models that we considered here are those
such that the total number of individuals is
constant in time. We do not consider demographic
processes, such as birth, death, and migration. 

We present the main features that are used to characterize
the spreading such as the reproduction number and the
frequency of new cases, which is called the epidemic curve
when it is plotted as a function of time. 
Our presentation emphasizes
the analogy with the thermodynamic phase transition,
the onset of spreading being understood as the
critical point of a continuous thermodynamic phase transition,
when the epidemic comes to an end, in which case the
frequency of new cases vanishes, the size of the epidemic
can be measured by the area under
the epidemic curve, and can be understood the order parameter.

The spatial stochastic version 
of the SIS model was proposed by Harris \cite{harris1974},
who named it the contact process. It is widely studied 
not only because it describes an epidemic spreading but also
because of its critical behavior \cite{grassberger1983a}
which is distinct from
models describing system in thermodynamic equilibrium such
as the Ising model \cite{tome2015L,marro1999,henkel2008}.
The SIS model describes the evolution of an infectious
disease that becomes endemic. The simplest model that 
describes the evolution of a infectious disease that
becomes extinct is the SIR model. Its spatial version
was formulated by Grassberger \cite{grassberger1983},
who called it general epidemic process, and its critical 
behavior is distinct from the Ising model and also different
from the SIS model \cite{tome2015L,marro1999,henkel2008}.

It is worth pointing out that the approach to epidemic
spreading, be it deterministic or stochastic, is
extended to a more general context of the population
dynamics \cite{tome2015L,ribeiro2017,cunha2017},
as for instance in ecological problems such as
the predator-prey dynamics \cite{satulovsky1994,tome2015L}.

\section{Evolution equations}

The spreading of an epidemic is understood as the time
evolution of a population consisting of several classes
of individuals pertaining to a certain community. 
The two main classes of individuals are the susceptible 
and the infective but other classes may also be present.
As the epidemic evolves in time, an individual belonging
in one class might change to another class through a
transforming process. We remark that the members of the
infective class are individuals that infect others.
An individual that is infected by a disease but does not
infect others belong in a distinct class, the exposed class.

The framework that is used here to analyze the 
population dynamics is borrowed from the theory of chemical kinetics.
According to this theory,
molecules of various species inside a vessel are subject to
chemical reactions that transform molecules of one
species into molecules of another species.
The several classes of individuals are analogous to the
chemical species, and the processes that transform an
individual of one class to an individual of another class are
analogous to chemical reactions. 

As an example of the analogy, let us consider the process
where an infective (I) individual recovers from the disease
becoming a recovered (R) individual. This process is 
represented by
\beq
{\rm I} \longrightarrow {\rm R},
\eeq
which is understood as a reaction that transforms one I
into one R, or the annihilation of one I and
the simultaneous creation of one R. This type of reaction
is called spontaneous.

Another example is the process of infection
of a susceptible (S) individual who become infected
by the contact with an infective (I) individual
It is represented by
\beq
{\rm S} \stackrel{\rm I}{\longrightarrow} {\rm I},
\label{r3}
\eeq
where the symbol I over the arrow means that the reaction
that transforms one S into one I needs the presence of one I.
This type of reaction is called catalytic or more precisely
auto-catalytic and I is the catalyst.

After introducing the analogy between chemical species
and chemical reactions, on one side, and classes of
individuals and transforming process, in the other,
we introduce a second analogy.
The time evolution of the numbers of individuals of each class,
determined by the transforming processes,
are understood as analogous to the 
chemical kinetic equations. These are ordinary 
differential equations of the first order in time involving
the concentrations or the fractions of the various chemical
species and are set up by the use of the law of mass action.

The primary question one wants to answer is how many
individuals are there in each class. Thus we
should seek for equations in the number of individuals
in each class. However, it is more convenient to
express the equations in terms of the density,
which is the number of individuals per unit area,
or in terms of the fraction of individuals in each 
class. To set up the evolution equations in either
one of these two types of variables, we will employ
the law of mass action, presented and explained
in the Appendix. Here, we choose to write down
the equations in terms of the fractions of each
class of individuals.

As an example of the application of the law of
mass action, consider the infection process represented
by equation (\ref{r3}),
and let us denote by $x$ the fraction of the susceptible
and by $y$ the fraction of the infective. The contribution
of this process to the evolution equation of both $x$ and $y$,
in accordance with the law of mass action, 
is $b x y$, where $b$ is the infection rate constant,
that is,
\beq
\frac{dx}{dt} = - b x y + \ldots,
\eeq
\beq
\frac{dy}{dt} = b x y + \ldots,
\eeq
where the dots indicate the contribution coming from
other processes. Notice the  presence of a minus sign
in the first equation, because the number of the susceptible
decreases in the infection process.

The evolution equations describing an epidemic
consist of two or more equations that give the
time variation of the fractions of the several classes of individuals.
We assume that the total number of individuals
is constant, so that the sum of all fractions equals one.
They are solved with an initial condition 
that reflects what occurs in a real epidemic.
At the beginning, all individuals are susceptible except
a very small fraction of them that are infective.

Our presentation uses frequently the term
rate, as in infection rate. It should not be confused with
the variation in time, which is a derivative. 
One should also make a distinction between a rate and
a rate constant which appears as a prefactor of a rate.

\section{Characterization}

\subsection{Epidemic curve}

A relevant characterization of the evolution of an epidemic is
the frequency of the number of individuals that are becoming
infected, or the number of newly infected individuals per unit
time, or simply the frequency of new cases. The ratio of
this number and the total number of individuals we denote
by $f$. When $f$ is plotted
as a function of the time it is called the epidemic curve.
The curve initially grows exponentially until an inflexion point
after which it 
behaves in such a way as to reach a final value which may be
zero, in which case the epidemic comes to an end,
or nonzero, in which case the disease becomes endemic.

To determine $f$, we consider all processes of the type
\beq
{\rm A} \stackrel{\rm I}{\longrightarrow} {\rm B},
\eeq
where B represents an infected and A a non-infected individual.
We then use
the law of mass action to determined the rate of
this process. The sum of the rates of all process of this
type is $f$. In the case of the auto-catalytic process of
the preceding section, in which an S is transformed
into one I, the frequency of new cases is $f=bxy$.

\subsection{Phase transition}

The spreading of a disease can be viewed as
a thermodynamic phase transition. To understand this
point we consider that the infection rate constant, 
denoted by $b$, takes several values. For small values,
there is no spreading whereas for large values
the epidemic spreads. Increasing $b$, one passes from
a non-spreading regime to a spreading regime at a critical
value $b_c$ that gives the onset of spreading. This condition
may seem to make no reference to the closeness of the individuals,
or equivalently to the density of the population. 
In a real process of infection, the contact, or 
at least the proximity of two individuals, is
an important condition for the onset of spreading.
Thus this feature should in some way be included in the theory.
In fact, this is the case, as explained next.

In the Appendix, it is shown that the infection rate constant $b$
is proportional to the intrinsic rate constant $b^*$, that is,
\beq
b = \gamma b^*,
\label{13}
\eeq
where $\gamma$ is the ratio of the total number of
individuals in a community and the area of the community,
or the population density. The intrinsic rate constant 
$b^*$ is a measure of the strength of infection
or the virulence of the disease and is density-independent.
Let us define the critical density $\gamma_c=b_c/b^*$
and see what happens when one increases $\gamma=b/b^*$.
When $\gamma$ increases, $b$ increases, and when
it reaches the value $\gamma_c$, the rate constant
reaches the critical value $b_c$, and
the spreading outbreaks. In other words, when the density
of the population reaches a critical value, the disease
outbreaks. This is, in fact, the statement of the Kermack and McKendrick
theorem, except that they refer to the density of the susceptible
but at the beginning of spreading the population consist
mostly of the susceptible.

It is usual to characterize the transition from a
trivial phase to the significant phase by an order
parameter, which is nonzero for the significant phase
and zero for the trivial one. For the cases of an
epidemic spreading that comes to an end, so that the
frequency of new cases $f$ vanishes for the long term,
the order parameter may be defined as the area $s$
of the epidemic curve,
\beq
s = \int_0^{\infty} f dt,
\eeq
where $f$ here is understood as function of $t$.

\begin{figure*}
\epsfig{file=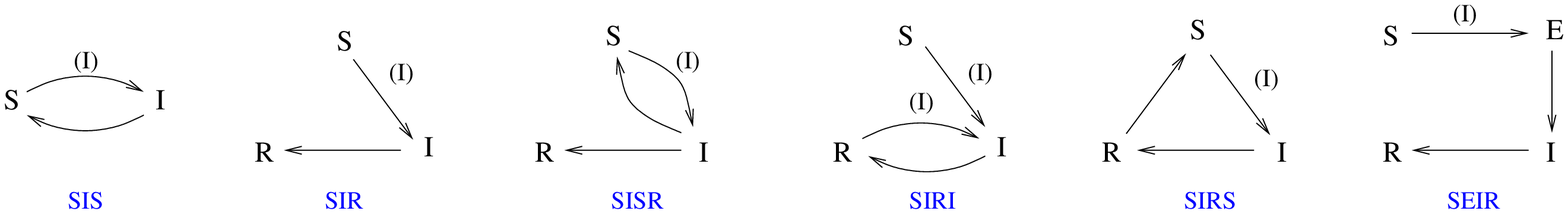,width=18cm}
\caption{Representation of the processes involved in each
epidemic spreading model. The classes of individuals are: susceptible (S),
infective (I), recovered (R), and exposed (E). 
An I between parentheses indicates that the process 
is catalytic and I is the catalyst. The other processes are spontaneous.
An exposed individual is infected but not infective.
A recovered individual has permanent immunity in the
SIR, SISR, and SEIR models.}
\label{diagr}
\end{figure*}

\subsection{Reproduction number}

Another important characterization of the epidemic spreading 
is related to the number of individuals 
that are infected by a single infective individual.
This number, called the reproduction number and denoted
by $R$, is the ratio of two numbers and is defined more
precisely as follows.
During an interval of time $\tau$, the number of new
cases is $N_1=fN\tau$, where $f$ is the frequency
of new cases defined above and $N$ is the total number
of individuals.

The question now arises about the number $N_2$ of the infective 
that are responsible for these new infections. If the number
of the infective remains the same in the interval $\tau$,
then $N_2$ would be equal to $N_1$. But if the infective
increases by a certain amount $N_3$ then $N_2$ should be smaller,
and in fact equal to $N_1-N_3$, and $R=N_1/(N_1-N_3)$. In view that
$N_3=N(dy/dt)\tau$, where $y$ is the fraction of infective
we find
\beq
R = \frac{f}{f-dy/dt}.
\label{1}
\eeq
As $f$ is the rate of a catalytic process, in which the infective
is the catalyst, $f$ is proportional to $y$, that is, $f=gy$,
which allows us to write
\beq
R = \frac{g}{g-d\ln y/dt}.
\label{1a}
\eeq

The basic reproduction number, denoted by $R_0$, is
the reproduction number at the early stages of
the spreading when the number of infective individuals
is negligible
and the whole population consists mostly of susceptible. 
Thus $R_0$ is obtained from $R$ by
setting $x=1$ in the formula ({\ref{1a}).

It is worth mentioning at this point a fundamental property 
that should be obeyed by the equations describing an epidemic.
As the spreading of the disease needs the presence of an
infected, then if $y=0$ there is no evolution meaning that
the other variables should not vary in time.
Therefore, the state consisting of $y=0$ and other
variables constant should be a solution of the evolution
equation. This state is the trivial solution and is
called disease-free solution. In stochastic models,
it is called absorbing state.

The stability analysis of the trivial solution is a way of
determining the outbreak of the spreading. Linear stability
reveals that the fraction of infective $y$ is dominated
by the exponential behavior, $y=y_0e^{\alpha t}$, where $\alpha$
is the largest eigenvalue of the stability matrix. The onset
of spreading occurs when $\alpha$ turns from a negative to
a positive value. Thus, at the early stages of the spreading
$y=y_0e^{\alpha t}$ and $x$ near one. Replacing
these results in equation (\ref{1a}) we reach a relationship
between the basic reproduction number and the exponent $\alpha$,
which is
\beq
R_0 = \frac{g_0}{g_0-\alpha},
\label{1b}
\eeq
where $g_0$ is the value of $g$ when $x=1$.
As the onset of spreading occurs when $\alpha=0$,
we see that the outbreak of spreading occurs at $R_0=1$,
When $\alpha>0$, that is, when $R_0>1$, the epidemic spreads.

\begin{figure*}
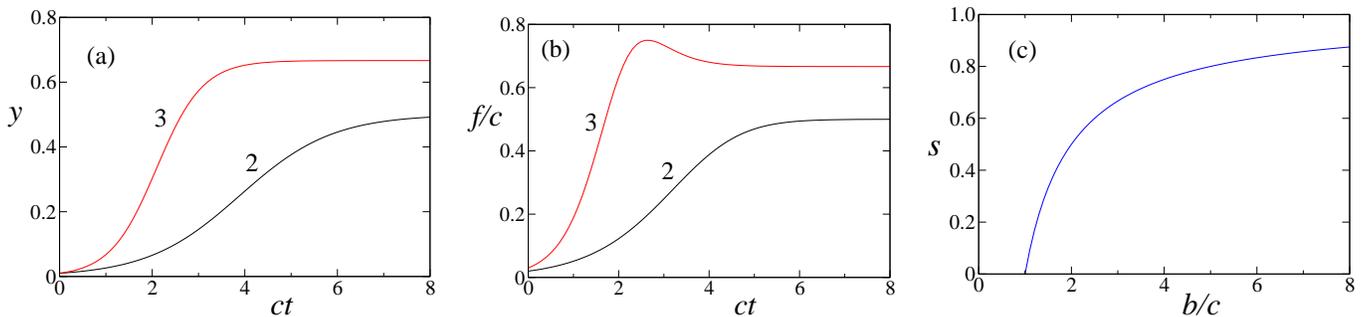

\epsfig{file=sis_inf.eps,width=5.7cm}
\hfill
\epsfig{file=sis_ec.eps,width=5.7cm}
\hfill
\epsfig{file=sis_op.eps,width=5.7cm}
\caption{SIS model.
(a) Fraction $y$ of infective individuals as a function of
time $t$, for the values of $b/c$ indicated. 
(b) Epidemic curve, which is the frequency $f$ of new cases
as a function of time $t$, for the values of $b/c$ indicated. 
(c) Order parameter $s$ versus the infection rate constant $b$. The order
parameter is the fraction of individuals that remain
infected for long times.}
\label{sis}
\end{figure*}

\section{Models of spreading}

The models that we will analyze consist of two or three processes
involving two or more classes of individuals. They are represented
by a flow diagram which gives all the possibilities of transformation
of the individual from one class to another. The six models that
we analyze are shown in figure \ref{diagr}.

\subsection{SIS model}

We start by considering the simplest model for
epidemic spreading which consists of two classes of individuals:
infective (I) and susceptible (S). The model comprises two
transformation processes. In the first, a susceptible individual becomes
infective by the contact with other infective individuals.
This process is represented by
\beq
{\rm S} \stackrel{\rm I}{\longrightarrow} {\rm I},
\label{r10}
\eeq
and occurs with an infection rate constant $b$.
In the second, an infective individual becomes susceptible spontaneously.
More specifically, an infective individual that has recovered
from the disease does not acquire immunity becoming 
immediately susceptible again. This process is represented by
\beq
{\rm I} \longrightarrow {\rm S},
\eeq
occurring with a recovery rate constant $c$.

We denote by $x$ and $y$ the fraction of the susceptible and
infective, respectively. As only these two classes
of individuals are present, the sum of these two fractions
equals one, $x+y=1$. According to the rules of mass action,
the equation that gives the evolution of $x$ and $y$ are
\beq
\frac{dx}{dt} = - bxy + cy,
\eeq 
\beq
\frac{dy}{dt} = bxy - cy.
\label{3}
\eeq 
Due to the constraint $x+y=1$, these equations are not independent.
Replacing $x$ by $1-y$ in equation (\ref{3}), it becomes
\beq
\frac{dy}{dt} = \alpha y - by^2,
\label{4}
\eeq 
where $\alpha=b-c$.

One solution of this equation is the trivial solution $y=0$,
and therefore $x=1$, which is understood as the absence of
epidemic spreading because there are no infective individuals
present. To find whether this solution is stable one tries to solve
the equation (\ref{4}) for small values of $y$. The term proportional
do $y^2$ is neglected and the equation becomes the linear equation
\beq
\frac{dy}{dt} = \alpha y,
\eeq 
whose solution is
\beq
y= y_0 e^{\alpha t},
\label{17}
\eeq
where $y_0$ is the value of $y$ at $t=0$.
Therefore, if $\alpha>0$, that is, if $b>c$, then
the number of infective increases exponentially
and the spreading of the disease sets in. If
$\alpha<0$, the disease does not spread. Recalling
that $\alpha=b-c$, it follows that $b=c$ marks the
onset of spreading. If $b>c$ the epidemic spreads
and if $b<c$ it does not.

The equation (\ref{4}) can in fact be solved in a
closed form by writing it in the differential form as
\beq
dt = \frac{dy}{y(\alpha-by)},
\eeq
or yet as
\beq
\alpha dt = \frac{dy}{y} + \frac{bdy}{\alpha-by}.
\eeq
Integrating we find
\beq
\alpha t = \ln y - \ln(\alpha-by) + k,
\eeq
where we are considering solutions such that $\alpha\geq by$.
Let us suppose that at $t=0$, the number of
infective is $y_0$. This determines the constant
$k$ and we may write
\beq
y = \frac{\alpha y_0}{by_0+(\alpha -by_0)e^{-\alpha t}},
\label{15}
\eeq
which gives $y$ as a function of $t$.

We distinguish two types of solutions depending on
the sign of $\alpha$. Let us consider that $\alpha<0$.
As we have seen above, in this case there is no spreading
of the disease. The right-hand side of equation (\ref{4})
is negative and $y$ will decrease toward its
asymptotic value $y^*$ which is found by taking the limit
$t\to\infty$ in equation (\ref{15}), which is $y^*=0$.
When $\alpha=0$, the exponential solution (\ref{15})
is no longer valid. To find the solution in this case,
one should solve the equation (\ref{4}) for $\alpha=0$,
which in this case reads
\beq
\frac{dy}{dt} = - by^2.
\label{4a}
\eeq 
The solution is
\beq
y = \frac{y_0}{by_0 t + 1}.
\eeq
For long times the decay towards the zero values is not
exponential but algebraic, $y\sim (bt)^{-1}$.

When $\alpha>0$, the
disease spreads and the fraction $y$ of the infective as
a function of time is shown in figure \ref{sis}a
for a small initial value of $y$.
The fraction of the infective
grows and then approaches and asymptotically the value $y^*$.
This asymptotic value is nonzero
and is found by taking the limit $t\to\infty$ in equation ({15}),
with the result
\beq
y^* = \frac{\alpha}{b} = \frac{b-c}{b}.
\label{7}
\eeq
Alternatively, $y^*$ can be found as the stationary
solution of equation (\ref{4}) which is obtained by setting to zero
the right-hand side of this equation. 

The frequency of new cases is the rate of the process (\ref{r10})
and thus given by
\beq
f = bxy = b(1-y)y.
\label{18}
\eeq
From the solution given by equation (\ref{15}), we may find
$f$ as a function of $t$, which is the epidemic curve
shown in figure \ref{sis}b.
We assume that initially the number of infective individuals
is small, which is the case of a real spreading of disease.

The initial increase of $f$ is exponential, a result that
reflects the initial exponential growth of $y$, given by the
equation (\ref{17}), that is,
\beq
f = b e^{\alpha t}.
\eeq
If $b/c\leq 2$, the epidemic curve $f$ increases monotonically
towards its final value $f^*=by^*(1-y^*)$ or
\beq
f^* = \frac{c}{b}(b-c).
\label{47}
\eeq
If $b/c > 2$, the epidemic curve increases initially, then
reaches a maximum and then decreases to its final value (\ref{47}).
In any of these cases the final value of $f^*$ is nonzero
which means that the disease becomes endemic.

The order parameter $s$ of the present model is
the final fraction $y^*$ of the infective, given by (\ref{7}),
\beq
s = \frac{b-c}{b}.
\label{7a}
\eeq
When $b\leq c$, $s=0$ and there is no spreading of the disease.
When $b>c$, the disease spreads. A plot of $s$ as a function
of $b$ is given in figure \ref{sis}c.

The reproduction number is given by equation (\ref{1}). 
As $f=bxy$ and $dy/dt$ is given by equation (\ref{3}) we find 
\beq
R = \frac{b}{c}x.
\eeq
The basic reproduction number is found by setting $x=1$,
\beq
R_0 = \frac{b}{c}.
\eeq

\begin{figure*}
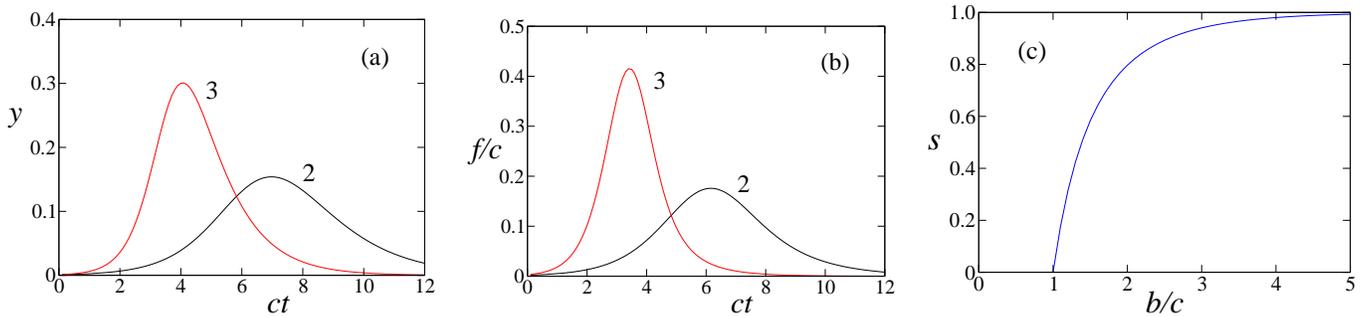

\epsfig{file=sir_inf.eps,width=5.7cm}
\hfill
\epsfig{file=sir_ec.eps,width=5.7cm}
\hfill
\epsfig{file=sir_op.eps,width=5.7cm}
\caption{SIR model. 
(a) Fraction $y$ of infective individuals as a function of
time $t$, for the values of $b/c$ indicated.
(b) Epidemic curve, which is the frequency $f$ of new cases
as a function of time $t$, for the values of $b/c$ indicated.
(c) Order parameter $s$ versus the infection rate constant $b$. The order
parameter is the area under the epidemic curve.}
\label{sir}
\end{figure*}

\subsection{SIR model}

This model differs from the SIS model in an important
feature. The infective individuals
after healing acquires permanent immunization,
meaning that they cannot be infected again. 
Therefore, in addition to susceptible (S) and infective (I) individuals,
there are the immune individuals, which we call recovered (R).
The model consists of two processes. The first
is the infection, represented by 
\beq
{\rm S} \stackrel{\rm I}{\longrightarrow} {\rm I},
\label{r7}
\eeq
occurring with an infection rate constant $b$,
and the second is the spontaneous recovery, represented by
\beq
{\rm I} \longrightarrow {\rm R},
\eeq
occurring with a recovery rate constant $c$,
which is strictly positive.

Denoting by $x$, $y$ and $z$, the number of 
susceptible, infective, and recovered individuals, respectively,
then according to the rule of mass action the evolution equation
of these variables are
\beq
\frac{dx}{dt} = - b xy ,
\label{8b}
\eeq
\beq
\frac{dy}{dt} = bxy - cy,
\label{8a}
\eeq 
\beq
\frac{dz}{dt} = cy.
\label{8c}
\eeq
As the total number of infective, susceptible and
recovered is constant, the fractions $x$, $y$, and $z$ are related
by $x+y+z=1$, and the three differential equations are not
all independent.

The differential equations (\ref{8b}), (\ref{8a}), and (\ref{8c})
were introduced by Kermack and McKendrick in their study of
epidemic spreading, allowing them to show the spreading threshold
theorem \cite{kermack1927}, which we show next.

One solution of the evolution equations is $x=1$, $y=0$ and $z=0$ which is 
the non-spreading state. To find the stability of this state,
we consider small deviations from this solution.
The equation for $y$ becomes
\beq
\frac{dy}{dt} =\alpha y,
\eeq
where $\alpha=b-c$, whose solution is
\beq
y=y_0 e^{\alpha t},
\label{21}
\eeq
and $y_0$ is the value of $y$ at $t=0$. Therefore,
if $\alpha>0$, that is, if $b>c$, the number of infective
grows exponentially and the spreading of the disease occurs.
If $\alpha<0$, $y$ decreases and the disease does not spread.
The onset of the spreading occurs when $b=c$.

To solve the time evolution equations we take the ratio between
equations (\ref{8c}) and (\ref{8b}) to reach the equation
\beq
\frac{dz}{dx} = - \frac{c}{bx},
\eeq
which can be integrated,
\beq
z = -\frac{c}{b}\ln x.
\label{11}
\eeq
The constant of integration was chosen by considering that
at the initial times, there is no recovered individuals, $z=0$,
and the number of infective is very small so that the
we may consider the fraction of the susceptible to be equal to one, $x=1$.
Replacing the result (\ref{11}) into $x+y+z=1$ we find
the following relation between $x$ and $y$,
\beq
y = 1-x + \frac{c}{b}\ln x,
\eeq
which replaced in (\ref{8b}) gives
\beq
\frac{dx}{dt} = - x (b - bx + c\ln x).
\eeq
In the integral form it reads
\beq
t = - \int \frac{dx}{x (b - bx + c\ln x)}.
\eeq

The integral can be solved numerically to get $x$ as
a function of $t$, from which we find $y$ and $z$.
Alternatively, these variables can be obtained by
solving numerically the 
set of equations (\ref{8b}), (\ref{8a}), and (\ref{8c}).
The result of $y$ as a function of $t$, for
a small initial value of $y$ and $z=0$, is shown
in figure \ref{sir}a. Initially there is an exponentially
growth, as determined by the equation (\ref{21}), then
$y$ reaches a maximum and then decreases  with time
and vanishes asymptotically.
This property is a consequence of the fact that
an infective individual that becomes healthy acquires
immunity and cannot be infected again. For long times
there will be no infective individuals but only the
recovered individuals and the ones that did not have
acquired the disease, the susceptible. 
The disease has become extinct.

The frequency of new cases is the rate of the process
(\ref{r7}) and given by
\beq
f = bxy,
\eeq
and for the SIR model it is also $f=-dx/dt$.
In figure \ref{sir}b we show $f$ as a function of time,
the epidemic curve. It was
obtained from the numerical solution of $x$ referred
to above. As initially the fraction of infective
is increasing exponentially so does the quantity $f$.
The epidemic curve increases, attains a maximum and
then decreases to the zero value. The area $s$ under the
epidemic curve, which is a measure of the size of the
epidemic, is the order parameter and will be determined next.

When $t\to\infty$, no infective is left, as we have already
seen, because once an infective individual is healed, it
acquires immunization and cannot be infected again. In other
words, the infective becomes recovered and remains in this
state forever. Thus when $t\to\infty$, $y\to0$, and the final
fraction of susceptible $x^*$ and the final fraction of 
recovered $z^*$ are related by $x^*+y^*=1$. 
Replacing $x^*=1-z^*$ in (\ref{11}) we find
\beq
z^* = -\frac{c}{b}\ln(1-z^*).
\label{49}
\eeq

We have seen above that $f=-dx/dt$.
In we integrate this equation in time, we find the order
parameter $s$, that is,
\beq
s = \int_0^\infty f dt = 1 - x^* = z^*,
\eeq
where we have taken into account that $x$ at the initial time
is equal to one. Replacing the result $z^*=s$ in equation
(\ref{49}) we find the equation that determines the order
parameter,
\beq
s = -\frac{c}{b}\ln(1-s).
\label{12}
\eeq

In figure \ref{sir}c we show $s$ as a function of the
infection rate constant $b$. When $b\leq c$, $s=0$ meaning that
the size of the epidemic is zero or that there is no
spreading of the disease. When $b>c$, the disease
spreads and the size of the disease increases as the
infection rate constant increases.
The equation (\ref{12}) is transcendental but can be solved
numerically. However, a solution can be written explicitly
when $s$ is small, in which case we find
\beq
s = 2\frac{b-c}{c}.
\eeq

The reproduction number is given by (\ref{1}). Using
$dy/dt$, given by equation (\ref{8a}), and recalling that 
$f=bxy$, we find 
\beq
R = \frac{b}{c}x.
\eeq
The basic reproduction number is found by setting $x=1$,
\beq
R_0 = \frac{b}{c}.
\eeq

\subsection{SISR model}

In the SIR model, whenever an infective individual
becomes healed it acquires permanent immunity.
Here we consider a modification of the SIR model
in which an infective individual may become again
susceptible \cite{tome2015L,ruziska2017}.
There are three processes. The first
is the infection, represented by
\beq
{\rm S} \stackrel{\rm I}{\longrightarrow} {\rm I},
\label{r12}
\eeq
occurring with an infection rate constant $b$.
The second is the spontaneous recovery, represented by
\beq
{\rm I} \longrightarrow {\rm R},
\eeq
occurring with a rate constant $c$, where R represents
an individual with permanent immunization. 
The third is the spontaneous recovery without immunization,
represented by
\beq
{\rm I} \longrightarrow {\rm S},
\eeq
occurring with a rate constant $a$. 
An individual recovers from the disease but does not
acquire immunization, becoming susceptible again.

The equation for the fractions $x$, $y$, and $z$
of susceptible, infective and recovered individuals are
\beq
\frac{dx}{dt} = - bxy + ay,
\label{23a}
\eeq
\beq
\frac{dy}{dt} = bxy - cy - ay,
\label{23b}
\eeq
\beq
\frac{dz}{dt} = cy,
\label{23c}
\eeq
and again $x+y+z=1$.

One solution of these equations is $x=1$, $y=0$, and $z=0$
corresponding to the absence of epidemic spreading.
For small values of $y$ and for $x$ near one, the equation
for $y$ reads
\beq
\frac{dy}{dt}=\alpha y,
\eeq
where $\alpha= b-c-a$. Therefore, if $b>c+a$ then the
epidemic spreading sets in whereas if $b<c+a$ there is no
spreading of the disease. 

The ratio of the equations (\ref{23c}) and (\ref{23a}) gives
\beq
\frac{dz}{dx} = - \frac{c}{bx - a},
\eeq
which can be integrated,
\beq
z = -\frac{c}{b}\ln\frac{bx-a}{b-a},
\label{64}
\eeq
with the condition $bx\geq a$.
The constant of integration was chosen by considering
that at the initial times, $z=0$, $y$ is negligible
so that we may take $y=0$ and $x=1$. 

The frequency of new cases $f$ is given by the
rate of the process (\ref{r12}) and is $f = b xy$.
The plot of $f$ as a function of $t$, the epidemic
curve, has a bell shape similar to that of the SIR model.
To get the area under this curve, which is the order
parameter $s$, we proceed as follows. 
By an appropriate combination of the evolution equations
(\ref{23a}) and (\ref{23c}), we write
\beq
bxy = - \frac{dx}{dt} + \frac{a}{c}\frac{dz}{dt},
\eeq
so that
\beq
f = - \frac{dx}{dt} + \frac{a}{c}\frac{dz}{dt}.
\eeq
Integrating $f$ in time from zero to infinity, we get
\beq
s = 1 - x^* + \frac{a}{c} z^*,
\label{55}
\eeq
where we recall that at $t=0$, $x=1$ and $z=0$,
and $x^*$ and $z^*$ are the final values of $x$ and $z$.

\begin{figure*}
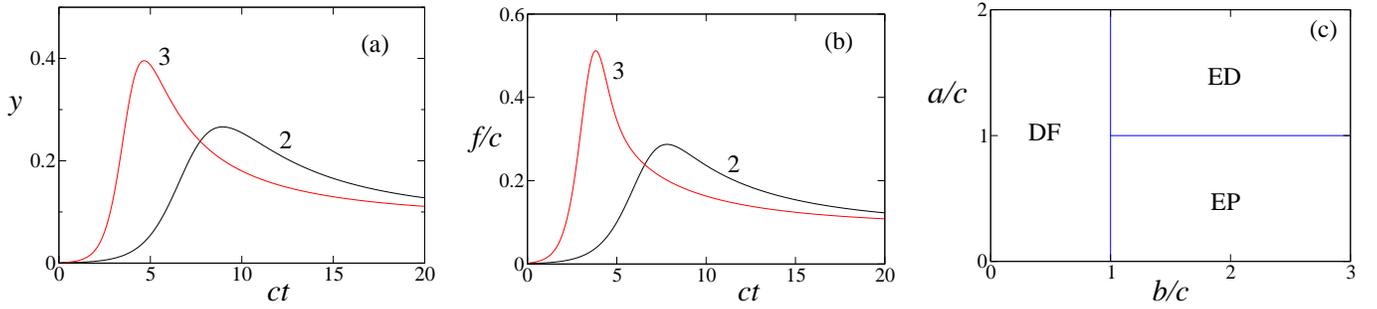

\epsfig{file=siri_inf.eps,width=5.7cm}
\hfill
\epsfig{file=siri_ec.eps,width=5.7cm}
\hfill
\epsfig{file=siri_diag.eps,width=5.7cm}
\caption{SIRI model.
(a) Fraction $y$ of infective individuals as a function 
of time $t$, for $a/c=1.1$ and the values of $b/c$ indicated.
The value of $y$ for long times is nonzero. 
(b) Epidemic curve, which is the frequency $f$ of new cases
as a function of time $t$, for $a/c=1.1$ and the values of $b/c$ indicated.
The asymptotic value of $f$ is nonzero.
(c) Phase diagram in the plane $a/c$ versus $b/c$ showing
the disease-free (DF), the endemic (ED) and the epidemic (EP) states.}
\label{siri}
\end{figure*}

To determine the value $z^*$,
we observe that for long times $y$ vanishes and $x^*=1-z^*$. 
Replacing this result in the equation (\ref{64}), we find
the equation for $z^*$,
\beq
z^* = -\frac{c}{b}\ln\left(1 - \frac{bz^*}{b-a}\right).
\label{14}
\eeq

Using the relation (\ref{55}) and the result $x^*=1-z^*$,
we find $s=(c+a)z^*/c$ and the size of the epidemic $s$
is proportional to the final value of the fraction
of the recovered $z^*$. 
The transcendental equation (\ref{14}) can be solved when 
$z^*$ is small, with the result
\beq
z^*=\frac{2}{b}(b-a-c).
\eeq

The threshold of the spreading of the disease occurs
when $b=a+c$. If $b<a+c$ there will be no spreading
and $s$ vanishes.
However, if $b>a+c$, the disease will spread and for long
times the fraction of the recovered individuals is $z^*$ given by
equation (\ref{14}).

The reproduction number is given by (\ref{1}). Replacing
$dy/dt$ given by equation (\ref{23b}) we find 
\beq
R = \frac{b}{c+a}x.
\eeq
The basic reproduction number is found by setting $x=1$,
\beq
R_0 = \frac{b}{c+a}.
\eeq

The numerical solution of the evolution equations
give results for the fraction $y$ and for the frequency
of new cases $f$ which are similar to those of the SIR
model. Both these quantities grow 
exponentially, attain a maximum value and then decrease 
to their final zero values. The order parameter has also a similar
behavior except that the critical point occurs when
$b=c+a$.

\subsection{SIRI model}

Here we consider another modification of the SIR model.
In the SIR model, an individual that has been infected
becomes immune, which means that a recovered individual remains
forever in this condition. In the present modification
the recovered individual loses the immunization and may
be reinfected again \cite{tome2015L,barros2017}. Therefore, to the
two processes of the SIR model,
\beq
{\rm S} \stackrel{\rm I}{\longrightarrow} {\rm I},
\label{r19}
\eeq
occurring with an infection rate constant $b$, and 
\beq
{\rm I} \longrightarrow {\rm R},
\eeq
occurring with a recovery rate constant $c$, we add the following process
\beq
{\rm R} \stackrel{\rm I}{\longrightarrow} {\rm I},
\label{r20}
\eeq
occurring with a reinfection rate constant $a$.
The equation for the fractions of susceptible, infective,
and recovered are
\beq
\frac{dx}{dt} = - b xy,
\label{13a}
\eeq
\beq
\frac{dy}{dt} = bxy - cy + ayz,
\label{13b}
\eeq 
\beq
\frac{dz}{dt} = cy - ayz,
\label{13c}
\eeq
and we recall that $x+y+z=1$.

The frequency of new cases are determined by the processes
(\ref{r19}) and (\ref{r20}) and is given by
\beq
f = bxy + ayz.
\label{65}
\eeq

\begin{figure*}
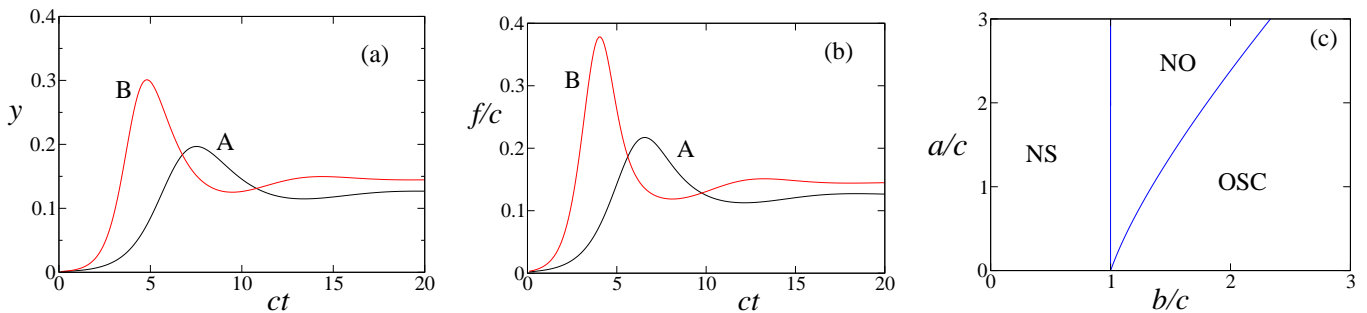

\epsfig{file=sirs_inf.eps,width=5.7cm}
\hfill
\epsfig{file=sirs_ec.eps,width=5.7cm}
\hfill
\epsfig{file=sirs_diag.eps,width=5.7cm}
\caption{SIRS model.
(a) Fraction $y$ of infective individuals as a function 
of time $t$. 
(b) Epidemic curve, which is the frequency $f$ of new cases
as a function of time $t$.
For the curve A: $a/c=1/3$ and $b/c=2$, and for the curve B:
$a/c=0.3$ and $b/c=2.7$.
(c) Phase diagram in the plane $a/c$ versus $b/c$ showing
the non spreading(NS), the oscillation (OSC) and non-oscillating
(NO) states.}
\label{sirs}
\end{figure*}

One solution of the evolution equations is the trivial
solution $x=1$, $y=0$, and $z=0$, which corresponds to
the absence of spreading. 
For small values of $y$ and for $x$ near one, the equation
for $y$ reads
\beq
\frac{dy}{dt}=\alpha y,
\eeq
where $\alpha= b-c$. Therefore, if $b>c$ the epidemic
spreads whereas if $b<c$ there is no spreading of the disease.

Dividing the equations (\ref{13c}) and (\ref{13a}) we find
\beq
\frac{dz}{c-az} = - \frac{dx}{bx},
\eeq
which can be integrated to give
\beq
\ln\frac{c-az}{c} = \frac{a}b \ln x,
\label{54}
\eeq
with the condition $az\leq c$. The constant of integration
was found by assuming as we did before 
that at the initial time, $z=0$ and $x=1$.

For long times we have two types of solution.
In one of them, the infected disappears, $y^*=0$,
and there remain the recovered, $z^*\neq 0$, and 
the susceptible, $x^*\neq0$. This solution occurs
when $a\leq c$ and the asymptotic value $z^*$ is obtained
by using (\ref{54}) and replacing
$x^*=1-z^*$. The result is the equation
\beq
\ln\frac{c-az^*}{c} = \frac{a}b \ln(1-z^*),
\eeq
which should be solved for $z^*$. For small values of $z^*$
the solution is
\beq
z^* = 2\frac{b-c}{c-a}.
\eeq
The time dependent solution is similar to the solution of the
SIR model.

As we have mentioned, $y^*=0$ for this solution so that
the frequency of new cases $f$ vanishes for long times
and the epidemic curve is also similar to that of the SIR model.
The area $s$ under the epidemic curve is given by the integral 
\beq
s = \int_0^\infty f dt,
\eeq
which can be obtained from the solution of the evolution equations.

Let us consider the other solution, which is valid for
$a>c$. The asymptotic values for this solution are
$x^*=0$, $y^*=(a-c)/a$, and $z^*=c/a$, which are obtained
by setting to zero the right-hand side of the evolution
equations (\ref{13a}), (\ref{13b}), and (\ref{13c}).
The time dependent solution leading to these values
can be obtained by solving numerically the evolution
equations. The fraction $y$ of infected is shown in
figure \ref{siri}a for the case $a/c=1.1$. 
From the solution for $x$, $y$ and $z$ we determine
the frequency of new cases using (\ref{65}), which
is shown in figure \ref{siri}b. This solution predicts
a persistence of the disease because $f$ remains finite
for long times.

The reproduction number is given by (\ref{1}). Replacing
$dy/dt$ given by equation (\ref{13b}) we find 
\beq
R = \frac{bx+az}{c}.
\eeq
The basic reproduction number is found by setting $x=1$,
in which case $z=0$, and
\beq
R_0 = \frac{b}{c}.
\eeq

\subsection{SIRS model}

We consider here a model with three classes of individuals
like the SIR model: susceptible, infective and recovered.
However, the present model consists of three instead of
two processes \cite{tome2015L,souza2010}. The first is
the infection of susceptible individuals, represented by
\beq
{\rm S} \stackrel{\rm I}{\longrightarrow} {\rm I},
\label{r25}
\eeq
occurring with an infection rate constant $b$.
The second is the spontaneous recovery, represented by
\beq
{\rm I} \longrightarrow {\rm R},
\eeq
occurring with a recovery rate constant $c$.
These two processes are the same as the SIR model.
In the present model, the recovered individuals are 
considered to have only partial immunity. Accordingly
they may become again susceptible through an 
spontaneous process represented by
\beq
{\rm R} \longrightarrow {\rm S},
\eeq
occurring with a rate constant $a$.

The evolution equations for $x$, $y$, and $z$, the fractions
of susceptible, infective and recovered individuals, are
\beq
\frac{dx}{dt} = - b xy + a z,
\label{15a}
\eeq
\beq
\frac{dy}{dt} = bxy - cy,
\label{15b}
\eeq 
\beq
\frac{dz}{dt} = cy - a z.
\label{15c}
\eeq
Again $x+y+z=1$ and the equations are not all independent.

The trivial solution is $x=1$, $y=0$, and
$z=0$, corresponding to the absence of spreading. 
Near this solution, that is,
for small values of $y$ and for $x$ near 1, the equation
for $y$ reads
\beq
\frac{dy}{dt}=\alpha y,
\eeq
where $\alpha= b-c$. Therefore, if $b>c$ then
$y$ increases exponentially and the spreading of the disease
occurs. If $b<c$, it does not.

To determine the  asymptotic fractions $x^*$, $y^*$, and
$z^*$ of each class of individuals at long term
we set to zero the right-hand sides of the equations
(\ref{15a}), (\ref{15b}), and (\ref{15c}).
The solution with a nonzero value of $y$ is
\beq
x^* = \frac{c}{b},
\qquad
y^* = \frac{a(b-c)}{b(a+c)},
\qquad
z^* = \frac{c(b-c)}{b(a+c)}.
\label{16}
\eeq

The frequency of new cases $f$ is determined by the process
(\ref{r25}) and is given by 
\beq
f = bxy.
\eeq
For long times it approaches a nonzero value given by
\beq
f^* = \frac{c^2(b-c)}{b(a+c)},
\eeq
and in this respect it is like the SIS model, that is,
the model predicts a persistence of the disease for long times,
or that the disease becomes endemic. However for
the present model the time dependence may present
damped oscillations as shown in figure \ref{sirs}.
This behavior is shown below.

Let us linearize the evolution equations (\ref{15a}) and (\ref{15b})
around the solution (\ref{16}). To this end, we define the
deviations $\xi=x-x^*$ and $\eta=y-y^*$ and write the
evolution equations in these new variables.
Up to linear terms in these variables, we find
\beq
\frac{d\xi}{dt} = M_{11}\xi + M_{12} \eta,
\eeq
\beq
\frac{dy}{dt} = M_{21} \xi + M_{22} \eta,
\eeq 
where $M_{11}=-a(a+b)/(a+c)$, $M_{12}=-(a+c)$,
$M_{21}=a(b-c)/(a+c)$, and $M_{22}=0$.
The stability of the solution above is determined
by the eigenvalues $\lambda$ of the matrix $M$ with elements $M_{ij}$
which are the roots of the equation
\beq
(a+c)\lambda^2 + a(a+b)\lambda + a(b-c)(a+c) = 0.
\eeq
The product of the roots is positive and the sum of them
is negative. If the roots are real then they are both
negative and the solution is stable. Now, let us
take a look at the complex roots which occurs when
\beq
a(a+b)^2\leq 4(b-c)(a+c)^2.
\label{67}
\eeq
In this case we should look at the real part of $\lambda$.
Since the real part is negative, then the solution is stable.

The conclusion is that the solution is stable but
$x$, $y$, and $z$, and also $f$, will display
damped time oscillations, as seen in figure \ref{sirs},
when the condition (\ref{67}) is fulfilled.
The values of $a$, $b$ and $c$ for this behavior is shown
in figure \ref{sirs}c.

\subsection{SEIR model}

In some infection diseases, the individual that has been
infected takes a certain time to be infective. To take into
account the latent period of such a disease, we introduce a class
of individual called exposed (E) who has been infected but is
not infective, but eventually becomes infective. A model that
includes the exposed individuals is similar to the SIR, but the
directed infection is replaced by the following two processes.
The first is represented by
\beq
{\rm S} \stackrel{\rm I}{\longrightarrow} {\rm E},
\label{r33}
\eeq
occurring with an infection rate constant $b$, and the
second is represented by
\beq
{\rm E} \longrightarrow {\rm I}
\eeq
occurring with a latent rate constant $h$, the inverse of which
is a measure of the latent period of the exposed individual.
The recover of an infective is the same as the SIR model,
represented by
\beq
{\rm I} \longrightarrow {\rm R},
\eeq
occurring with a recovery rate constant $c$.
The evolution equations for the fractions $x$, $u$, $y$, and $z$,
of susceptible, exposed, infective and recovered, respectively, are
\beq
\frac{dx}{dt} = - bxy,
\label{20a}
\eeq
\beq
\frac{du}{dt} = bxy - h u,
\label{20b}
\eeq
\beq
\frac{dy}{dt} = h u - c y,
\label{20c} 
\eeq
\beq
\frac{dz}{dt} = cy.
\label{20d}
\eeq
These equations are not all independent because $x+u+y+z=1$.
These equations were introduced by Dietz to account for
the latent period of infection \cite{dietz1976}.

The frequency of new cases $f$ is determined by the process
(\ref{r33}) and is given by
\beq
f = bxy.
\eeq
Considering that $bxy=-dx/dt$, it follows that the area
under the epidemic curve is
\beq
s = \int_0^\infty f dt = 1 - x^*,
\eeq
where we have considered that at the initial times $x=1$.

The trivial solution of the equations (\ref{20a}), (\ref{20b}),
(\ref{20c}), and (\ref{20d}) is $y=0$, $u=0$
and $x+z=1$. To determine the stability of this solution,
we linearize the equation around the solution $y=0$, $u=0$,
$z=0$ and $x=1$. The equation for $u$ and $y$ become
\beq
\frac{du}{dt} = b y - h u,
\eeq
\beq
\frac{dy}{dt} = h u - c y.
\eeq
To find the solution of these equations
we assume that
\beq
u=u_0 e^{\alpha t} \qquad y=y_0 e^{\alpha t}.
\label{53}
\eeq
Replacing this solution in the equations of evolution we find
\beq
- h u_0 + b y_0 = \alpha u_0,
\label{54a}
\eeq
\beq
h u_0 - c y_0 = \alpha y_0,
\label{54b}
\eeq
which are understood as a set of eigenvalue equations
where $\alpha$ is understood as the eigenvalue.
The possible values for $\alpha$ are the roots of
\beq
\alpha^2 + (h+c)\alpha + h(c-b) = 0,
\eeq
the largest one being
\beq
\alpha = \frac12\{ -(h+c) + \sqrt{(h-c)^2+4hb} \}.
\label{27}
\eeq
If $\alpha<0$, which occurs when $b<c$, the disease does
not spread. If $\alpha>0$, which occurs when $b>c$, the
disease spreads. In this regime, $y$ and $u$ grow
exponentially with an exponent $\alpha$ given by equation (\ref{27}).
Let us compare this exponent with that of the SIR model, which is
\beq
\alpha_{\rm sir} = b-c.
\eeq
In both cases the onset of spreading occurs when $b=c$.
However, when $b>c$, that is, in the spreading regime, 
$\alpha<\alpha_{\rm sir}$ indicating that
the epidemic growth rate is smaller for the 
SEIR model than it is for the SIR model. This is due
to the necessary passage of an individual to the
exposed state before reaching the infective state.

The behavior of the present model concerning the
epidemic curve and the time behavior of fraction of infective 
is qualitatively similar to those of the SIR model
shown in figure \ref{sir}. The behavior of the
size $s$ of the epidemic is the same as that of
the SIR model and does not depend on the rate constant $h$
as we show in the following. 

\begin{figure}
\epsfig{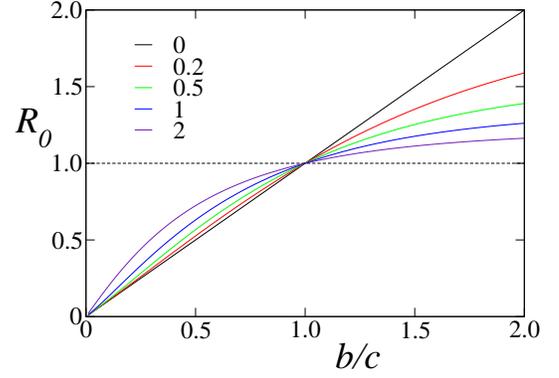}
\caption{Basic reproduction number $R_0$ for the SEIR model
as a function of the reduced infection rate $b/c$ for several
values of the latent period $\ell=c/h$ of an exposed individual. 
Notice that, the SIR model corresponds to $\ell=0$.
The outbreak of the epidemic occurs when $R_0=1$.}
\label{rzero}
\end{figure}

We divide the equation (\ref{20d}) by the equation (\ref{20a})
\beq
\frac{dz}{dx} = - \frac{c}{bx},
\eeq
which can be solved to give
\beq
z = - \frac{c}{b}\ln x,
\eeq
where the constant of integration was found by
considering that at the early stages, $z=0$ and $x=1$.
For long times the infective as well as the exposed
disappear, $y=0$ and $u=0$. Therefore, the sum of
the final values $x^*$ and $z^*$ of the fractions
of the susceptible and of the recovered equals one
and $x^*=1-z^*$. Recalling that the order parameter
$s=1-x^*=z^*$ we reach the following equation for $s$,
\beq
s = -\frac{c}{b}\ln (1-s)
\eeq
which is identical to the equation (\ref{12}) for
the SIR model. It is worth mentioning that this equation
says that $s$ does not depend on $h$ which means that
the size of the epidemic is independent of $h$. In
other terms, the presence of the exposed does not
change the size of the epidemic. This process only
slows the velocity of the spreading but not its size
which is the same as that of the SIR model.

The reproduction number is given by (\ref{1}). Replacing
$dy/dt$ given by equation (\ref{20c}) we find 
\beq
R = \frac{bx}{bx-h(u/y)+c}.
\eeq
To find the basic reproduction number we need to know the
ratio $u/y$ when $x=1$, that is, in the early stages of
the spreading. According to (\ref{53}), this ratio equals $u_0/y_0$
so that 
\beq
R_0 = \frac{b}{b-h(u_0/y_0)+c}.
\eeq
The ratio can be found from the eigenvalue equation (\ref{54b}).
Dividing this equation by $y_0$ we find
\beq
h \frac{u_0}{y_0} - c = \alpha,
\label{54c}
\eeq
and $R_0$ acquires the form
\beq
R_0 = \frac{b}{b-\alpha},
\eeq
where $\alpha$ is the eigenvalue (\ref{27}).
A plot of $R_0$ is given in the figure \ref{rzero}
for various values of the latent period $\ell=c/h$.
Notice that, the SIR corresponds to the case when
the exposed rapidly becomes infective, that is,
when the latent period is zero.

\section{The Ross theory of happenings}

Ronald Ross received the Nobel Prize for Medicine in 1902 for his
work on the transmission of malaria. He believed that it was not
necessary to eliminate all mosquitoes to prevent the spreading
of the disease but only to reduce the density of mosquitoes 
to a value below the critical level \cite{heesterbeek2015}.
This conception as well as that of the threshold theorem of
Kermack and McKendrick are understood as instances of the
fundamental result of the statistical mechanics of interacting
systems that the transition from one regime to another
occurs only when one reaches a critical value of a parameter. 

Ross advanced the idea of applying differential calculus
to the dynamics of an infectious disease with the aim
of its control and prevention. He even coined the name
{\it pathometry} for what we call theoretical epidemiology.
The theory introduced in his 1911 book on the prevention of
malaria he called theory of happenings \cite{ross1911}. 
It is worth mentioning that the book contained the reports
on the condition of malaria in several regions of the world,
written by several authors. This includes the report
by Oswaldo Cruz, the Brazilian epidemiologist, on the
campaign on the prevention of malaria in southwest Brazil
carried out in the first decade of the twentieth century.

His theory of happenings on the spreading of an infectious disease,
Ross presented in an appendix to his book on the prevention of 
malaria of 1911 \cite{ross1911} and in subsequent papers 
\cite{ross1915,ross1916}, some of which
with Hilda Hudson \cite{ross1917a,ross1917b}.
In a population of $P$ individuals affected by a certain disease, 
the time variation $dZ/dt$ in the number of the affected $Z$
is $hA+qZ$ where $A=P-Z$ is the number of unaffected, 
$h$ is the proportion of unaffected that becomes affected
and $q$ takes into account the demographic and recovery rates.
For infectious diseases, Ross argues that $h$ is proportional
to the fraction $x=Z/P$ of the affected, and writes $h=cx$
where $c$ is a constant which Ross calls the infection rate,
and arrives at the following equation
\cite{ross1911,ross1915,ross1916}
\beq
\frac{dx}{dt} = cx(\ell-x).
\label{2}
\eeq
The solution of this equation, given by Ross, is
\beq
x = \frac{x_0 \ell}{x_0 + (\ell-x_0)e^{-c\ell t}},
\label{97}
\eeq
revealing that $x$ increases slowly at the beginning and
then very rapidly until it reaches an inflexion, 
and after that approaches the limiting value $\ell$
\cite{ross1915,ross1916}.
According to Ross the current proportion of new cases
to the total population is \cite{ross1915,ross1916}
\beq
f = c x (1-x).
\label{98}
\eeq

The equation (\ref{2}) and its solution (\ref{97})
are identical to the equations (\ref{4}) and (\ref{15}),
respectively, and the frequency of new cases (\ref{98})
is identical to $f$ given by (\ref{18}). 
These results shows that the model considered by Ross 
is identified with the SIS model that we have analyzed.

\section{Conclusion}

We have presented an analysis of deterministic models for epidemic
spreading. The equations were set up by using the analogy 
between chemical reactions and the processes occurring
in the epidemic spreading which are understood as a
change of the class of an individual. The main analogy
was the use of the law of mass action, which provides the
rate of the several processes that define an epidemic model.
We have also emphasized the analogy of the onset of 
an epidemic with a thermodynamic phase transition.
When the infection constant changes it reaches a critical
value at which the spreading takes place. The infection
constant is argued to depend on the density of the population
and as it increases the spreading outbreaks at a certain
critical density, which is the theorem advanced by
Kermack and McKendrick.

We have also analyzed the quantities that characterizes the
spreading of an epidemic. One of them is the frequency of
new cases, which is the number of new infected individuals
per unit time. When this quantity vanishes the epidemic
comes to an end. In this case the area of the epidemic curve
is a measure of the epidemic and 
may be identified as the order parameter. It may happen
that the frequency of new cases does not vanishes and in this
case the disease becomes endemic. We have seen that the
SIRS model which is appropriate for this case, predict 
oscillations in the frequency of new cases.

Another way of determining the outbreak of the spreading
is by means of the stability analysis of the disease-free
state which is the state without any infective individuals.
Any model of spreading must have this state which in
stochastic models is called absorbing state. 
The stability analysis gives the behavior of the spreading
at the beginning and shows that the growth
of the epidemic is exponential. We have related the growth exponent
with the basic reproduction number. When the
exponent change signs, which means that the reproduction
number passes from a value less than one to a value greater
than one, the spreading outbreaks.

All the properties were obtained by solving the evolutions
equations, which are ordinary differential equations of
first order in time. This may be obtained in closed form
or by numerical methods.
We have finally showed that the SIS model was in fact the model
originally studies by Ross in his theory of happenings.

\appendix

\section{Law of mass action}

The law of mass action \cite{tome2015L} was formulated originally by 
Guldberg and Waage in the context of the equilibrium of chemical reactions
\cite{waage1864a,waage1864b,guldberg1864}.
Since then it was used in the chemical kinetic theory
and in other fields such as that of the present application.

Let us consider a set of chemical reaction represented by
\beq
\sum_{i({\rm reac.})} \nu_{ij} A_i\,\, 
\to \!\! \sum_{i({\rm prod.})} \nu_{ij}' A_i,
\eeq
where $A_i$ represents a molecule of the $i$ species and
$\nu_{ij}\geq 0$ and $\nu_{ij}'\geq 0$ are the stoichiometric
coefficients related to the $j$-th reaction.
The sum on the left-hand side is over the reactants whereas
that on the right is over the products of the reaction.
A catalytic reaction is understood as a reaction where
a certain molecule $A_i$ appears both as a reactant and as a
product of the reaction,
with the same stoichiometric coefficient, $\nu_i'=\nu_i>0$,
and an auto-catalytic, if $\nu_i'>\nu_i>0$.

Let us denote by $\rho_i$ the concentration of molecules of species $A_i$,
which is the number of molecules per unit volume. 
The evolution equation for $\rho_i$ is postulated to be of the form
\beq
\frac{d\rho_i}{dt} = \sum_j (\nu_{ij}'-\nu_{ij}) g_j,
\label{42}
\eeq
which is a sum of terms, one for each one of the reactions
in which the molecule $A_i$ appears as a reactant or as a 
product of the reaction. According to the law of mass action,
the term $g_j$ is given by
\beq
g_j = k_j^* \prod_i (\rho_i)^{\nu_{ij}},
\eeq
where $k_j^*$ is the intrinsic reaction rate constant 
related to the $j$-th reaction, and the mathematical
product in $i$ is over the {\it reactants}.

Suppose we wish to formulate the equations above
in terms of the fraction $x_i$ of each species. If we 
denote by $N$ the total number of molecules then the
relation between density and fractions is $ N x_i = V \rho_i$
because each one of these two terms equals the number
of molecules of species $i$. Defining $\gamma=N/V$
this relation can be written as
\beq
\rho_i = \gamma x_i.
\label{41}
\eeq
From now on we consider that $N$, or equivalently
$\gamma$, is invariant in time, or that it varies so
slowly compared to the variation of the fractions
$x_i$, that it might be taken as constant in time.
With this condition in mind, we replace (\ref{41})
in (\ref{42}) to find the
evolution equation in terms of the fractions,
\beq
\frac{dx_i}{dt} = \sum_j (\nu_{ij}'-\nu_{ij})f_j,
\eeq
where $f_j$ is 
\beq
f_j = k_j \prod_i (x_i)^{\nu_{ij}},
\eeq
and the mathematical product in $i$ is again over the {\it reactants}.
The reaction rate constant $k_j$ is proportional to the 
intrinsic reaction rate constant $k_j^*$ and related to
$\gamma$ by
\beq
k_j = k_j^* \gamma^{-1} \prod_i \gamma^{\,\nu_{ij}}.
\label{58}
\eeq

Let us consider the spontaneous reaction
\beq
{\rm A}_1 \longrightarrow {\rm A}_2.
\eeq
The rate constant $k$ of this reaction is identical to the
intrinsic rate constance $k^*$, that is, $k=k^*$.
However, for the auto-catalytic reaction
\beq
{\rm A}_1 + {\rm A}_2 \longrightarrow {\rm A}_1 + {\rm A}_3,
\eeq
the rate constant $k$, according to the formula (\ref{58}),
is $\gamma$ times the intrinsic rate constant $k^*$, that is,
$k=\gamma k^*$.

The distinction between a rate constant and an intrinsic
rate constant is that the intrinsic is free from concentration
dependence whereas the rate constant may depend
on the concentration as shown in the second example above.


\end{document}